\font\eightrm=cmr8 at 8pt
\documentclass{oxarticle}
\usepackage{amsmath, amssymb}
\usepackage{graphics, setspace}
\usepackage{epstopdf}
\usepackage{bm}
\usepackage{tikz}\usepackage{xcolor}

  \newcommand{\cdss}{chiral dotted spinor superfield}

\newcommand{\XM}{{Exotic Model}}

\newcommand{\ME}{Master Equation}

\newcommand{\cM}{{\cal M}}

\newcommand{\CDSS}{{Chiral Dotted Spinor Superfield}}

\newcommand{\cA}{{\cal A}}

\newcommand{\Exists}{\bm{\exists}\kern-0.6em\bm{\exists}}

\newcommand{\EI}{Exotic Invariant}
\newcommand{\ei}{exotic invariant}

\newcommand{\SSM}{Supersymmetric Standard Model}

\newcommand{\cI}{{\cal I}}

\newcommand{\bt}{\begin{tabular}{c}}
\newcommand{\et}{\end{tabular}}

\newcommand{\eb}{\ee\be } 
\newcommand{\ebp}{\rt.\ee\be\lt.}

\newcommand{\bmat}{\lt ( \begin{array} }
\newcommand{\emat}{  \end{array} \rt )}

\newcommand{\oX}{{\ov X}}

\newcommand{\oa}{{\ov a}}

\newcommand{\oY}{{\ov Y}}

\newcommand{\oy}{{\ov \y}}

\newcommand{\oC}{{\ov C}}

\newcommand{\oF}{{\ov F}}
\newcommand{\A}{{\ov A}}

\renewcommand{\a}{\alpha}	
\renewcommand{\b}{\beta}
\newcommand{\g}{\gamma}
\renewcommand{\d}{\delta}

\newcommand{\m}{\mu}
	
\newcommand{\x}{\xi}

\newcommand{\f}{\phi}
\renewcommand{\c}{\chi}
\newcommand{\y}{\psi}

\newcommand{\G}{\Gamma}

\newcommand{\Lam}{\Lambda}

\renewcommand{\S}{\Sigma}

\newcommand{\la}{\label}
\newcommand{\ci}{\cite}

\newcommand{\ds}{\documentstyle}	
\newcommand{\fr}{\frac}

\newcommand{\pa}{\partial}
\newcommand{\ov}{\overline}
\newcommand{\br}{\begin{rant}}
\newcommand{\er}{\end{rant}}
\newcommand{\beC}{\begin{Conjecture}}
\newcommand{\eeC}{\end{Conjecture}}
\newcommand{\be}{\begin{equation}}
\newcommand{\ee}{\end{equation}}
\newcommand{\ba}{\begin{array}} 
\newcommand{\ea}{\end{array}}
\newcommand{\bea}{\begin{eqnarray}}
\newcommand{\eea}{\end{eqnarray}}

\newcommand{\Ra}{\Rightarrow}

\newcommand{\lt}{\left}
\newcommand{\rt}{\right}

\newcommand{\ben}{\begin{enumerate}}
\newcommand{\een}{\end{enumerate}}
\newcommand{\bitem}{\begin{itemize}}
\newcommand{\eitem}{\end{itemize}}

\newcounter{orange} 
\newcounter{apple} 
\addtocounter{apple}{1}
\newcounter{grape} 
\addtocounter{grape}{1}
\newcommand{\numberhere}{Feb1\;}
\newcommand{\articlenumber}{E3-FINAL.tex}

\proofmodefalse
\usepackage{color}

\newcommand{\mathsym}[1]{{}}
\newcommand{\unicode}[1]{{}}

\begin{document}
 
 \begin{center}

{
\
\
\;{\Large The  simplest Exotic Invariant  (E3)
\\[1cm]}  
%

\renewcommand{\thefootnote}{\fnsymbol{footnote}}

{\Large John A. Dixon\footnote{jadixg@gmail.com, john.dixon@ucalgary.ca}\\Physics Dept\\University of Calgary \\[1cm]}  }
\end{center}

\normalsize
\Large
 
 \begin{center}Abstract
\end{center}

  This paper E3 shows how to construct the simplest Exotic Invariant in the simplest way.

\section{Introduction}

Interesting examples of the BRS cohomology of SUSY exist. They are called ``Exotic Invariants".  In this paper E3, the simplest exotic invariant is constructed in the simplest way.

 These \ei s are important  for  the \SSM.  This paper is the third paper E3 in  the E series of papers on Exotic Invariants. There will soon be ten of these papers \ci{E1}...\ci{E??}.

\section{The  Action $\cA$ for the  theory with no superpotential}
\la{actionsection} 

We will start with the action for the simplest chiral supersymmetry theory.  But we will need to formulate it in the context of the BRS results, and we also need to add some extra pieces.  The action   contains a number of subactions.  The main two pieces are:
\be
\cA= \cA_{\rm Fields} +\cA_{\rm PseudoFields} 
\la{sumfofieldandpseudo}
\ee
where  the part $\cA_{\rm Fields}$ in 
(\ref{sumfofieldandpseudo}) is as follows:

\subsection{Field parts $\cA_{\rm Fields}$ of the action}

This part in turn consists of a number of parts:
\be
\cA_{\rm Fields}= {\cal A}_{\rm A\;Kinetic} +\cA_{\rm CDSS,\;Kinetic}  + \cA_{\rm CDSS, \;Chiral\;Kinetic}  +\ov\cA_{\rm CDSS, \;Chiral\;Kinetic}
\la{susyaction}
\ee
These are as follows.  First we have the usual kinetic action for a chiral multiplet:
\be {\cal A}_{\rm A\;Kinetic} = 
a_0 \int d^{4}x \left \{
  F    {\ov F}
  -
\y^{ \a  }  \pa_{ \a \dot \a  }  {\ov \y}^{  \dot \a}  
-  \pa_{ \m  }
  A   \pa^{ \m  }  \A 
\rt\}
\la{Akineticaction}\ee

$A$ is a scalar field, $\y^{ \a  }$ is a spinor field, and $F$ is an auxiliary scalar field. We include dimensionless constants like $a_0,a_1 \cdots $, because these coefficients will be very important. We do not want mass terms and interactions here for the simplest case.  Next we have the, quite unusual, but simple, kinetic action for a \cdss\ (``CDSS") .  It consists of three parts.  The \cdss\ has two different possible kinetic terms--this is very important.  Here they are:

 \be
\cA_{\rm CDSS,\;Kinetic}    
= a_1 \int d^4 x   
\lt \{  {\f}_{\dot \b} \pa^{\a \dot \b} \Box {\ov \f}_{\a} 
+ X_{\a \dot \b} \pa^{\a \dot \d} \pa^{\g \dot \b}   {\ov X}_{ \g \dot \d }  
+  {\c}_{\dot \b} \pa^{\a \dot \b}  {\ov \c}_{\a}  
\rt \} 
\la{CDSSkineticaction}
\ee
\be \cA_{\rm CDSS, \;Chiral\;Kinetic}  =  a_2  \int d^4 x   
\lt \{2  \f^{\dot \b}   \Box      \c_{\dot \b}  
 +    X^{\a \dot \b }  \Box      X_{\a\dot \b }
\rt \} 
\la{CDSSchiralkineticaction}\ee

\be  \ov\cA_{\rm CDSS, \;Chiral\;Kinetic} =  \oa_2  \int d^4 x   
\lt \{2  {\ov \f}^{ \b}  \Box   {\ov \c}_{ \b}  
 +   {\ov  X}^{ \b \dot \a } \Box      {\ov  X}_{ \b  \dot \a  }
\rt \} 
\la{CDSSchiralkineticactionbar}\ee 
$X_{\a \dot \b}$ is a complex vector field, ${\f}_{\dot \b}$ is a spinor field, and ${\c}_{\dot \b}$ is another spinor field. 
We do not include mass terms and interactions here for the simplest case.
\subsection{PseudoField parts $\cA_{\rm PseudoFields}$ of the action}
The part $\cA_{\rm PseudoFields}$ in 
(\ref{sumfofieldandpseudo}) is as follows:

\be
\cA_{\rm PseudoFields}= {\cal A}_{\rm PseudoFields,A}  
+\cA_{\rm  PseudoFields,CDSS}+ * \eb + {\cal A}_{\rm Structure}
\la{pseudoaction}
 \ee
The following term is a standard type of term for the formulation of the BRS Master Equation in the simplest chiral theory.  It consists of source terms, which we call pseudofields, coupled to the BRS variations of the various fields in (\ref{Akineticaction}):
\be
{\cal A}_{\rm PseudoFields, A}  = \int d^4 x \;
\lt\{ \G     \lt ( \y_{ \b} {  C}^{  \b} 
+ \x^{\a \dot \b} \partial_{\a \dot \b}  A \rt )
\la{psdueoA1}\ebp +
 Y^{ \a}
 \lt ( \pa_{ \a \dot \a  }  A \oC^{\dot\a}
 + F {C}_{ \a}
+ \x^{\nu} \partial_{\nu}  \y_{  \a}
\rt )+
\Lam_{E}  \lt ( {\ov C}^{\dot \a}
 \pa_{ \a \dot \a  }  {  \y}^{  \a}
 + \x^{\nu} \partial_{\nu}   F 
\rt )
\rt \}
\la{psdueoA2}\ee
The above (\ref{psdueoA1}) , (\ref{psdueoA2}) could also be written 
\be
{\cal A}_{\rm PseudoFields, A}  = \int d^4 x \;
\lt\{ \G     \lt ( \d A \rt )
 +
 Y^{ \a}
 \lt ( \d \y_{  \a}
\rt )+
\Lam_{E}  \lt ( \d   F 
\rt )
\rt \}
\ee

The variations of the \cdss\ fields are quite similar:
\be
{\cal A}_{\rm PseudoFields, CDSS} = \int d^4 x \;
\lt \{ G^{\dot\a}\lt (  C^{\a} X_{\a \dot \a}+
 \x^{\g \dot \d} \partial_{\g \dot \d}\f_{\dot \a}\rt )
\la{CDSSpseudo1} 
 \ebp
+\S^{ \dot \b \a}\lt (
\pa_{ \a \dot \g }  \f_{\dot\b} {\ov C}^{\dot \g}  
+ 
C_{\a}   
\c_{\dot\b}
+ \x^{\g \dot \d} \partial_{\g \dot \d}  X_{\a \dot \b}
\rt )
+L^{\dot \a}\lt (
  \pa^{\b \dot \g}  X_{ \dot \a \b}   {\ov C}_{\dot \g} 
+ \x^{\g \dot \d} \partial_{\g \dot \d}  \c_{\dot\a}
\rt ) \rt \}
\la{CDSSpseudo2} 
 \ee
The above (\ref{CDSSpseudo1}) , (\ref{CDSSpseudo2}) could also be written 
\be
{\cal A}_{\rm PseudoFields, CDSS} = \int d^4 x \;
\lt \{ G^{\dot\a}\lt (  \d \f_{\dot \a}\rt )
+\S^{ \a\dot \b } \lt (\d X_{\a \dot \b}
\rt )
+L^{\dot \a}\lt (\d  \c_{\dot\a}
\rt ) \rt \}
 \ee
Finally there is the structure constant term:
\be
 {\cal A}_{\rm Structure} =
 -K_{\a \dot \b} C^{\a} \oC^{\dot \b}
\ee
This is more important than it seems--it is the origin of the all the BRS cohomology and the \ei s, as was shown in \ci{E2}.

\section{The  Simplest Example }
\la{simpexamp}
We need to show how the simplest example is invariant. Here is the simplest
 example\footnote{ This simple example comes from the following item in the spectral sequence:
\be
(C \x^2 C) \f^{\dot \a} \oy_{\dot \a}\Ra \cI_1 
\ee
which is discussed elsewhere in these papers.}:
\be
\cI_1 = \int d^4 x \lt \{ e_1  \f^{\dot \a} \G \oC_{\dot \a} +e_2 
 X^{\b \dot \a}Y_{\b}\oC_{ \dot\a}  +e_3 
 \c^{\dot \a}\Lam \oC_{ \dot\a}  \ebp
 +e_4
  X^{\b \dot \a}\pa_{\b \dot\a} \A+e_5 
 \c^{\dot \a}\oy_{ \dot\a} 
\rt \}
\la{simpexampeq} 
\ee

The constants $e_r, r=1\cdots 5$ are  determined by requiring invariance: 
\be
\d \cI_1 =0
\ee
The operator $\d$ is defined in the following two sections, and we can get it from them.

Let us see how this works for the first term
\be
\d  \cI_1=
\int d^4 x \lt \{ e_1 \d \f^{\dot \a} \G \oC_{\dot \a} \rt \}
-\int d^4 x \lt \{ e_1\f^{\dot \a} \d  \G \oC_{\dot \a} \rt \}+ e_2 \cdots
\eb= 
- \int d^4 x \lt \{ e_1 \lt (   - C_{\a} X^{\dot \a \a}+ \x^{\g \dot \d} \partial_{\g \dot \d}   \f^{\dot\a}
\rt ) \G \oC_{\dot \a} \rt \}
\la{linedssfsdf}\eb
-\int d^4 x \lt \{ e_1\f^{\dot \a} \lt (  \Box    {\ov  A}_{} 
-\pa_{ \a \dot \b } Y^{ \a}    {\ov C}^{\dot \b}   
+ \x^{\g \dot \d} \partial_{\g \dot \d} \G\rt ) \oC_{\dot \a} \rt \}+ e_2 \cdots
\ee
Consider the following term from line (\ref{linedssfsdf}):
\be
+\int d^4 x \lt \{ e_1  C_{\a} X^{\dot \a \a} \oC_{\dot \a} 
 \G \rt \}
\la{firstbit}\ee
What in the variation could this cancel against? We can gaze at the variations in section \ref{brsoperatorsection2} for a while and then we see that 
\be
\d  \int d^4 x \lt \{ e_2 
 X^{\b \dot \a}Y_{\b}\oC_{ \dot\a} \rt \}
=\eb
 \int d^4 x \lt \{ e_2 
\d  X^{\b \dot \a}Y_{\b}\oC_{ \dot\a} \rt \}
+ \int d^4 x \lt \{ e_2 
   X^{\b \dot \a} \d Y_{\b}\oC_{ \dot\a} \rt \}
\ee
contains the relevant piece as follows:
\be
 \int d^4 x \lt \{ e_2 
   X_{\a \dot \a} \d Y^{\a}\oC^{ \dot\a} \rt \}
\ee
\be
=
\int d^4 x \lt \{ e_2 
   X_{\a \dot \a}\lt ( 
-
  \pa^{\a \dot \b  }   
{\ov \y}_{ \dot \b}
-
\G  
 {C}^{  \a}
-
  \pa^{\a \dot \b}  \Lam  {\ov C}_{\dot \b} 
  + \x^{\g \dot \d} \partial_{\g \dot \d}  Y^{ \a}
\rt )\oC^{ \dot\a} \rt \}
\ee
The second term in the above is the one we want for now.  It is
\be
- \int d^4 x \lt \{ e_2 
   X_{\a \dot \a} 
\G  
 {C}^{  \a}
\oC^{ \dot\a} \rt \}
\la{secondbit}\ee

So now we can combine the two pieces (\ref{firstbit}) and (\ref{secondbit}) as follows:
\be
+\int d^4 x \lt \{ e_1  C_{\a} X^{\dot \a \a} \oC_{\dot \a} 
 \G \rt \}
- \int d^4 x \lt \{ e_2 
   X_{\a \dot \a} 
\G  
 {C}^{  \a}
\oC^{ \dot\a} \rt \} =0
\ee
and this implies 
\be
 e_1 - e_2 =0
\la{e1ise2}
\ee
The rest of the determination of these $e_i$ works in a similar way.  We are assured that a solution exists, with nonzero coefficients $e_i$, because the spectral sequence guarantees it. Deriving the rest of the coefficients is a similar task--it requires one to remember that $\d$ anticommutes with Grassmann odd objects like $\G,\Lam$ and spinors, and one needs to juggle indices quite a lot, and remember that $C$ is Grassmann even and that $C$  does not transform. It is also essential to remember identities such as:
\be
C^{\a} \y_{\a} = -C_{\a} \y^{\a} 
\ee

 If one actually goes through this exercise, the content of the theory becomes much clearer, of course. Are there errors? Always! Are there a very large number of important minus signs?  Yes. Is it a dull and boring task?  Certainly!  That is why the only hope for this kind of theory is to get it done by computer, and we will return to that in \ci{E10}.

\section{The \ME}
\la{mastereqsection}

We shall not try to derive the following here.  The literature on Master Equations in the context of BRS is huge.  See for example \ci{pigsib} to \ci{dixmin}.   We will simply assert that the following is true for this theory.  The  $\cA$ here is given above in equation (\ref{sumfofieldandpseudo}). The \ME\ has the following form.
\be
\cM_{\rm }=\cM_{\rm A} 
+\cM_{\rm X}+ \cM_{\rm\A} 
+\cM_{\rm \oX}  +\cM_{\rm Structure}=0
\ee
where
\be
\cM_{\rm A}=
\int d^4 x \lt \{
\fr{\d \cA}{\d A} \fr{\d \cA}{\d \G}
+
\fr{\d \cA}{\d \y_{ \a}  } \fr{\d \cA}{\d Y^{\a}}
+
\fr{\d \cA}{\d F} \fr{\d \cA}{\d \Lam } \rt \}
\ee

\be
 \cM_{\rm \A}=
\lt \{\fr{\d \cA}{\d \A} \fr{\d \cA}{\d \ov\G }
 +
\fr{\d \cA}{\d \oy_{ \dot \a}} \fr{\d \cA}{\d \oY^{\dot \a}}
+
\fr{\d \cA}{\d \oF } \fr{\d \cA}{\d \ov\Lam }
\rt \}  
\la{wzh}
\ee

\be
\cM_{\rm X} =\int d^4 x \lt \{
\fr{\d \cA}{\d \f_{\dot\a}} \fr{\d \cA}{\d  G^{\dot\a}}
+
\fr{\d \cA}{\d X_{\a\dot\a}} \fr{\d \cA}{\d \S^{\a\dot\a}}
+\fr{\d \cA}{\d\c_{\dot\a}} \fr{\d \cA}{\d L^{\dot\a}}
 \rt \}
\ee

\be
\cM_{\rm \oX} =\int d^4 x \lt \{
\fr{\d \cA}{\d \ov \f_{\a}} \fr{\d \cA}{\d  \ov G^{ \a}}
+
\fr{\d \cA}{\d \ov X_{\dot\a\a}} \fr{\d \cA}{\d \ov\S^{\dot\a\a}}
+
\fr{\d \cA}{\d\ov\c_{\a}} \fr{\d \cA}{\d \ov L^{\a}}
\rt \}  
\la{mastercdssdownfield} 
\ee
\be
 \cM_{\rm Structure}=
\fr{\pa \cA}{\pa K_{\a \dot \b}} 
\fr{\pa \cA}{\pa \x^{\a \dot \b}}    
\la{masterstructure}
\ee

This Master equation summarizes the invariance of the action (\ref{susyaction}) under the supersymmetry transformations that arise from the parts (\ref{pseudoaction}) and the \ME\ also includes the closure of the algebra of supersymmetry.

 \section{Tables of the Nilpotent BRS operator $\d$  these are older I think}
 \la{brsoperatorsection2}
 
 By taking the ``square root'' of the Master equation we can derive the following ``nilpotent BRS transformations'' of the fields and pseudofields above.
\be
\la{brstransE} 
\vspace{.1in}
\framebox{{$\begin{array}{lll}  
& &{\rm Nilpotent  \;Transformations\; for\; the \; A\;Fields}\\
\d A&= & 
\fr{\d {\cal A}}{\d \G} 
=  \y_{  \b} {C}^{  \b} 
+ \x^{\g \dot \d} \partial_{\g \dot \d} A
\\
\d {\ov A} &= & 
\fr{\d {\cal A}}{\d {\ov \G}} 
=  {\ov \y}_{ \dot \b} {\ov C}^{ \dot  \b} 
+ \x^{\g \dot \d} \partial_{\g \dot \d} {\ov A}
\\

\d \y_{ \a} &  =& \fr{\d {\cal A}}{\d {  Y}^{   \a} } = 
\pa_{ \a \dot \b } A {\ov C}^{\dot \b}  
+ 
C_{\a}   
F
+ \x^{\g \dot \d} \partial_{\g \dot \d}  \y_{\a  }
\\

\d
 {\ov \y}_{ \dot \a} &  =& 
\fr{\d {\cal A}}{\d { {\ov Y}}^{i\dot   \a} } = 
\pa_{ \a \dot \a }  {\ov A}_{} {C}^{\a}  
+ 
{\ov C}_{\dot \a}   
{\ov F}
+ \x^{\g \dot \d} \partial_{\g \dot \d} 
 {\ov \y}_{ \dot \a} 
\\
 \d F 
&=&\fr{\d {\cal A}}{\d \Lam} = 
  \pa_{\a \dot \b}   \y^{ \a} {\ov C}^{\dot \b} 
+ \x^{\g \dot \d} \partial_{\g \dot \d}  F 
\\
 \d \oF 
&=&\fr{\d {\cal A}}{\d \ov\Lam} = 
  \pa_{\b \dot \a}   \oy^{\dot \a} { C}^{\b} 
+ \x^{\g \dot \d} \partial_{\g \dot \d}  \oF 
\\
\end{array}$}} 
\ee

\be
\la{brstransEPs} 
\vspace{.1in}
\framebox{{$\begin{array}{lll}  
& &{\rm Nilpotent  \;Transformations\; for\; the \; A \; pseudofields}
\\
\d \G 
&= &
 \fr{\d {\cal A}}{\d A} 
=
  \Box    {\ov  A}_{} 
-\pa_{ \a \dot \b } Y^{ \a}    {\ov C}^{\dot \b}   
+ \x^{\g \dot \d} \partial_{\g \dot \d} \G
\\
\d {\ov \G} 
&= & \fr{\d {\cal A}}{\d {\ov A}} 
=
 \Box          { A}   -
 \pa_{ \a \dot \b } {\ov Y}_E^{ i \dot \b}    {C}^{\a}   
+ \x^{\g \dot \d} \partial_{\g \dot \d} 
 {\ov \G_E}
\\
\d Y^{ \a} 
&=&\fr{\d {\cal A}}{\d {  \y}_{  \a}} 
= 
-
  \pa^{\a \dot \b  }   
{\ov \y}_{ \dot \b}
-
\G  
 {C}^{  \a}
-
  \pa^{\a \dot \b}  \Lam  {\ov C}_{\dot \b} 
  + \x^{\g \dot \d} \partial_{\g \dot \d}  Y^{ \a}
\\
\d 
{\ov Y}^{ \dot \a} 
&=&\fr{\d {\cal A}}{\d {\ov \y}^{ \dot \a} 
} 
= 
-
  \pa^{\b \dot \a  }   
{ \y}_{ \b}
-
{\ov \G} 
 {\ov C}^{\dot  \a}
- 
  \pa^{\b \dot \a}  \ov\Lam   { C}_{\b} 
+ \x^{\g \dot \d} \partial_{\g \dot \d}  
{\ov Y}^{\dot \a} 
\\
 \d \Lam  
&=&  \fr{\d {\cal A}}{\d {F} } =
\oF 
+ Y^{\a}
  C_{\a}+ 
\x^{\g \dot \d} \partial_{\g \dot \d}   \Lam  
\\
 \d {\ov \Lam}  
&=&  \fr{\d {\cal A}}{\d {\ov F} } =
F
+ \oY^{ \dot \a}
  \oC_{\dot\a}+ 
\x^{\g \dot \d} \partial_{\g \dot \d}   {\ov \Lam}   
\\
\end{array}$}} 
\ee

\be
\vspace{.1in}
\framebox{{$\begin{array}{lll}  
 & &{\rm Nilpotent \; CDSS\;Transformations}\\
\d \f_{\dot\a}&= & 
\fr{\d {\cal A}}{\d G^{\dot\a}} 
=  (C^{\d} X_{\dot \a \d}) + \x^{\g \dot \d} \partial_{\g \dot \d}   \f_{\dot\a}

 \\
\d \f^{\dot\a}&= & 
\fr{\d {\cal A}}{\d G_{\dot\d}} 
=  (- C_{\a} X^{\dot \a \a})+ \x^{\g \dot \d} \partial_{\g \dot \d}   \f^{\dot\a}
 \\
\d X_{\a \dot \b}  &  =& 
\fr{\d {\cal A}}{\d  \S^{\a \dot \b } } = 
(\pa_{ \a \dot \d }  \f_{\dot\b} {\ov C}^{\dot \d}  
+ 
C_{\a}   
\c_{\dot\b})+ \x^{\g \dot \d} \partial_{\g \dot \d}  X_{\a \dot \b} 
 \\
\d X^{\b \dot \a}  &  =& 
\fr{\d {\cal A}}{\d  \S_{\b\dot \a} } = 
(-\pa^{ \b \dot \d }  \f^{\dot\a} {\ov C}_{\dot \d}  
+ 
C^{\b}   
\c^{\dot\a})+ \x^{\g \dot \d} \partial_{\g \dot \d}   X^{\b \dot \a} 
 \\
 \d \c_{\dot\a}
&=&\fr{\d {\cal A}}{\d L^{\dot \a}} = 
(  \pa^{\d\dot \d}  X_{\d\dot \a}   {\ov C}_{\dot \d} )+ \x^{\g \dot \d} \partial_{\g \dot \d}   \c_{\dot\a}
 \\
 \d \c^{\dot\a} 
&=&\fr{\d {\cal A}}{\d L_{\dot \a}} = 
(  \pa_{\d \dot \d}  X^{\d \dot \a}   {\ov C}^{\dot \d} )+ \x^{\g \dot \d} \partial_{\g \dot \d}   \c^{\dot\a}
 \\
\end{array}$}} 
\ee

\be
\vspace{.1in}
\framebox{{$\begin{array}{lll}  
 & &{\rm Nilpotent \; CDSS\;Pseudofield\;Transformations}\\
\d G^{\dot\a}&= & 
\fr{\d {\cal A}}{\d \f_{\dot\a}} 
=  {\rm etc. \;are\; not\; needed\;here\;for\;now}
\\
\end{array}$}} 
\ee

\be
{\cal A}_{\rm ZJStructure} =  -K^{\a \dot \b}    C_{\a} {\ov C}_{ \dot \b}
\ee

\section{Conclusion}

In this paper E2, we have provided the bare minimum necessary for a reader to verify that the expression (\ref{simpexampeq}) is invariant under the transformations in section \ref{brsoperatorsection2}.  The essential   argument is in section \ref{simpexamp}, which results in the simple equation (\ref{e1ise2}).

\begin{center}
 { Acknowledgments}
\end{center}
\vspace{.1cm}

  I thank     Friedemann Brandt, Philip Candelas,   Mike Duff, Sergio Ferrara,  Dylan Harries, Marc Henneaux,  D.R.T. Jones, Olivier Piguet, Antoine van Proeyen,     Peter West and Ed Witten for stimulating correspondence and conversations.   I also express appreciation for help in the past from William Deans, Lochlainn O'Raifeartaigh, Graham Ross, Raymond Stora, Steven Weinberg, Julius Wess and Bruno Zumino. They are not replaceable and they are missed.  I also  thank Ben 
Allanach, Doug Baxter,  Margaret Blair,  Murray Campbell, David Cornwell, Thom Curtright, James Dodd, Richard Golding, Chris T.  Hill,  Davide Rovere,   Pierre Ramond, Peter Scharbach,  Mahdi Shamsei, Sean Stotyn, Xerxes Tata and J.C. Taylor, for recent, and helpful, encouragement to carry on with this work. I also express appreciation to Dylan Harries and  to Will, Dave and Peter Dixon, Vanessa McAdam and Sarunas Verner for encouraging and teaching me to use coding.  I note with sadness the recent passing of Carlo Becchi and Kelly Stelle, both of whom were  valuable colleagues and good friends.


\tiny 
\articlenumber\\
\today
\hourandminute

\end{document}